\documentclass[superscriptaddress,prl,twocolumn,amsmath,amssymb]{revtex4}
\usepackage{graphicx}
\usepackage{dcolumn}
\usepackage{bm}
\usepackage[british]{babel}
\usepackage{array} 

\newcommand{\be}{\begin{equation}}
\newcommand{\ee}{\end{equation}}
\newcommand{\ba}{\begin{eqnarray}}
\newcommand{\ea}{\end{eqnarray}}
\newcommand{\ban}{\begin{eqnarray*}}
\newcommand{\ean}{\end{eqnarray*}}

\newcommand{\moy}[1]{\langle #1 \rangle}

\newcommand{\ket}[1]{\mbox{$ | #1 \rangle $}}
\newcommand{\bra}[1]{\mbox{$ \langle #1 | $}}

\newcommand{\demi}{\frac{1}{2}}

\begin{document}

\title{Reference frame independent quantum key distribution}

\author{Anthony Laing}
\email{anthony.laing@bristol.ac.uk}
\affiliation{Centre for Quantum Photonics, H. H. Wills Physics Laboratory \& Department of Electrical and Electronic Engineering, University of Bristol, BS8 1UB, United Kingdom}
\author{Valerio Scarani}
\email{ physv@nus.edu.sg}
\affiliation{Centre for Quantum Technologies and Department of Physics, National University of Singapore, Singapore}
\author{John G. Rarity}
\email{john.rarity@bristol.ac.uk}
\affiliation{Centre for Quantum Photonics, H. H. Wills Physics Laboratory \& Department of Electrical and Electronic Engineering, University of Bristol, BS8 1UB, United Kingdom}
\author{Jeremy L. O'Brien}
\email{jeremy.obrien@bristol.ac.uk}
\affiliation{Centre for Quantum Photonics, H. H. Wills Physics Laboratory \& Department of Electrical and Electronic Engineering, University of Bristol, BS8 1UB, United Kingdom}

\date{\today}

\begin{abstract}
We describe a quantum key distribution protocol based on pairs of entangled qubits that generates a secure key between two partners in an environment of unknown and slowly varying reference frame.  A direction of particle delivery is required, but the phases between the computational basis states need not be known or fixed.  The protocol can simplify the operation of existing setups and has immediate applications to emerging scenarios such as earth-to-satellite links and the use of integrated photonic waveguides. We compute the asymptotic secret key rate for a two-qubit source, which coincides with the rate of the six-state protocol for white noise. We give the generalization of the protocol to higher-dimensional systems and detail a scheme for physical implementation in the $3$ dimensional qutrit case.
\end{abstract}

\maketitle

\textit{Introduction. ---} Technologies based on the principles of quantum information \cite{nielsen} promise a revolution in informational tasks such as computer processing \cite{sh-conf-94-124, de-prsla-400-97} and communication \cite{be-prl-69-2881}. Secure communication via quantum key distribution (QKD) is one quantum information application that can be realized with current technologies \cite{gi-rmp-74-145, du-poi-4-463, sc-rmp-81-1301, lo-arxiv-2008}. All the photonic QKD protocols proposed to date have in common the need for a \textit{shared reference frame} between the authorized partners Alice and Bob: alignment of polarization states for polarization encoding, interferometric stability for phase encoding. This requirement can in principle be dispensed with by encoding logical qubits in larger-dimensional many-photon physical systems \cite{boileau}.  However, the creation, manipulation and detection of many-photon entangled states, are both technically challenging and very sensitive to the losses on the Alice-Bob channel --- in a word, impractical. To date, therefore, all practical implementations of QKD within an environment of varying phase, have required the frames of Alice and Bob to be actively aligned by classical communication.

\begin{figure}[h!]
\includegraphics[trim=40 320 230 140, clip, width=8cm]{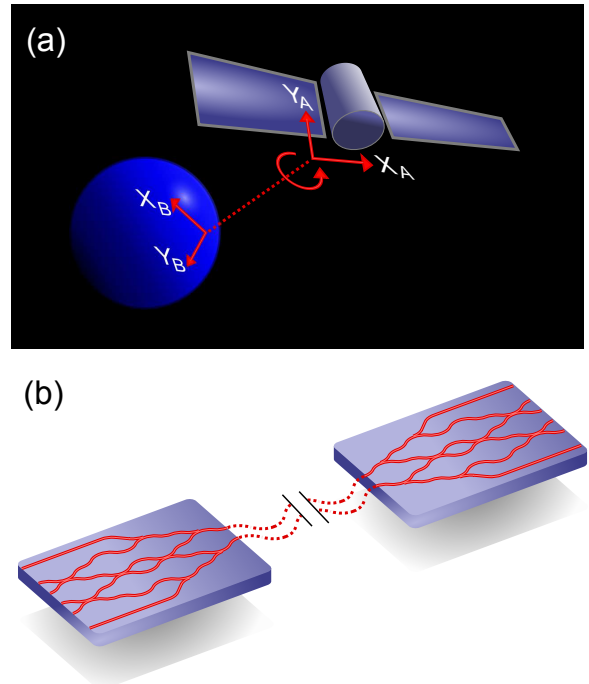}
\vspace{-4mm}
\caption{(Color online) Two meaningful scenarios for reference frame independent QKD. (1) Polarization encoding in earth-to-satellite quantum communication.  Here, the circular polarisation states are stable, but the linear states can vary with the rotation of the satellite.  (2) Path encoding in chip-to-chip quantum communication.  While the path information is stable, the unpredictable wavelength-scale changes in relative path length amount to a varying reference frame.  This may occur between chips communicating through free space, or between chips connected by optical fibres.}
\vspace{-5mm}
\label{figscenarios}
\end{figure}

In this paper, we present a \textit{reference frame independent (rfi) protocol} that can be implemented with ordinary sources and operate without frame alignment, beyond the obvious establishment of a particle delivery link. Moreover, there are at least two emerging scenarios in QKD that will benefit from an rfi implementation (Figure \ref{figscenarios}). The first such scenario is \textit{earth-to-satellite QKD} \cite{ra-njp-4-82,ku-nat-419-450,pe-prl-94-150501,sp-oc-260-340, rfiQKDnote1, Mi-pla-361-29,ur-nphys-3-481,sc-prl-98-010504,bo-njp-11-045017}. In this case, one axis of the reference frame is well defined: the beam must obviously connect the earth station with the satellite. On this beam, information encoded in circular polarization is very stable, but the linear polarizations may vary in time because the satellite may be rotating with respect to the ground station.  The second scenario is path encoded \textit{chip-to-chip QKD}.  The monolithic structures of planar waveguides have been successfully used to perform the stable interferometric measurements required in time and phase encoded QKD \cite{ma-leo-3-589, ho-opletts-29-2797, ta-pra-72-041804, fu-apl-95-261103}.  More recently integrated quantum photonic circuits have demonstrated their potential as components for more general quantum information tasks \cite{po-sci-320-646, matthews-2008, po-sci-325, ma-oe-17-12546, sm-oe-17-13516, la-hifi}.  In these latter cases path encoding is typically used, enabling deterministic single photon manipulations, in contrast to the probabilistic manipulations used in time bin encoding.  In a path encoded chip-to-chip setup, the ``which path'' information is very stable, but it is unthinkable to expect interferometric stability between two separate channels connecting the Alice and Bob chips. In these and similar scenarios, our protocol leads to the generation of a secure key without aligning the frames, as long as the repetition rate of the signals is faster than the rate of change of frame. 

\textit{The protocol for two qubits. ---} For ease of notation, we denote by $\{X,Y,Z\}$ the three Pauli matrices usually written $\{\sigma_x,\sigma_y,\sigma_z\}$. We assume that one direction is well defined, which is the case for all the usual encodings in QKD: the circular basis in polarization encoding, the time basis in time-bin encoding, the which-path basis in path encoding. So we set $Z_A=Z_B$. The other two directions are related by $X_B=\cos\beta X_A+\sin\beta Y_A$ and $Y_B=\cos\beta Y_A-\sin\beta X_A$, where $\beta$ may vary in time.

We present the protocol in its entanglement-based version where Alice and Bob share the state $\rho_{AB}$, which in the ideal case is the $|\phi^{+}\rangle$ Bell state; an equivalent prepare-and-measure version can be obtained through the usual recipe (see e.g. II.B.2 in \cite{sc-rmp-81-1301})

In each run, Alice and Bob choose independently one of the three directions (randomly but not necessarily with the same probability) and measure the quantum signal they receive in the corresponding basis. At the end of the signal exchange, they reveal their bases. The \textit{raw key} consists of the cases where both have measured in the $Z$ basis; so the quantum bit error rate is given by
\ba
Q&=&\frac{1-\moy{Z_AZ_B}}{2}\,.
\ea 
In order to estimate Eve's knowledge, Alice and Bob need to use the information collected on the bases complementary to $Z$. The quantity
\ba
C&=&\moy{X_AX_B}^2+\moy{X_AY_B}^2\\ &+& \moy{Y_AX_B}^2+\moy{Y_AY_B}^2
\ea
is independent of the relative angle $\beta$ and will be used to bound Eve's knowledge. The maximal value under Pauli algebra is $C=2$, achievable only by (a subset of) two-qubit maximally entangled states --- note that, in this case, one has $Q=0$ as well: the two parameters $C$ and $Q$ are not independent, as we shall see in more detail later.

Before turning to a formal security proof, it is important to understand how the protocol is affected by the fact that $\beta$ may vary in time. $C$ being a statistical quantity, its estimation requires several repetitions of the experiment. A variation of $\beta$ during the run will have the effect of smearing the estimated correlations.  For the protocol to be useful, therefore, Alice and Bob should collect sufficient signals to create a key above the finite-size effects \cite{scaren,caisca}, in a time short enough for $\beta$ not to vary too much. Now, while the expected variations of $\beta$ should be estimated in order to assess the feasibility of an implementation, during the run of the protocol $\beta$ is not a parameter available to Alice and Bob: its monitoring would amount to aligning the frames, which defeats the purpose. In the context of security assessment, any smearing of the correlations will be attributed to Eve's intervention.

\textit{Security bound. ---} As we have just seen, since $\beta$ is not monitored by Alice and Bob, we have to assume the worst case scenario: $\beta$ is fixed, known to Eve, and all the smearing of the correlations is due to Eve's intervention. We derive an asymptotic security bound against coherent attacks by an eavesdropper, under the assumption that the source produces a two-qubit state. 

As a \textit{first step}, we notice that Alice and Bob process each pair independently of the others. This fact, together with the assumption that we are dealing with finite-dimensional systems, guarantees that we can compute the bound by restricting to collective attacks \cite{kgr,rgk}. Thus, each pair shared by Alice and Bob is supposed to be in the two-qubit state $\rho_{AB}$, of which Eve holds a purification.

The \textit{second step} consists in proving that we can consider $\rho_{AB}$ (or just $\rho$ for ease of notation) to be Bell-diagonal in some Bell-basis known to Eve, without loss of generality. The proof is similar to the one presented in Refs \cite{devindep,devindepfull}. First, we use the fact that $C$ is invariant under the transformation $X_A\rightarrow-X_A$, $Y_A\rightarrow-Y_A$, $X_B\rightarrow-X_B$ and  $Y_B\rightarrow-Y_B$. This transformation can be implemented on $\rho$ itself as the unitary $Z_AZ_B$. In the presence of such a symmetry, it is not restrictive to replace $\rho$ by $\tilde{\rho}=\demi\big(\rho+Z_AZ_B\rho Z_AZ_B\big)$: indeed, if Eve can gain some knowledge out of $\rho$, she can gain the same knowledge out of $Z_AZ_B\rho Z_AZ_B$; by mixing them, she can therefore gain at least the same knowledge, and maybe more because the state is more mixed. As for Alice and Bob, they do not notice any difference, since they are looking \textit{only} at $Q$ and $C$. So presently we have
\ba
\tilde{\rho}_{AB}&=&\mu_1P_{\Phi^+}+\mu_2P_{\Phi^-}\,+\,\Big(\frac{a}{2}\ket{\Phi^-}\bra{\Phi^+}+H.c.\Big)
\nonumber\\
&+&\mu_3P_{\Psi^+}+\mu_4P_{\Psi^-}\,+\,\Big(\frac{b}{2}\ket{\Psi^-}\bra{\Psi^+}+H.c.\Big)
\ea
where $P_{\psi}=\ket{\psi}\bra{\psi}$ and the four states represent the usual Bell basis. For convenience of notation, let us call this state $\tilde{\rho}(a,b)$. Now, we have $C=2\left[(\mu_1-\mu_2)^2+(\mu_3-\mu_4)^2+ \mathrm{Im}(a)^2+\mathrm{Im}(b)^2\right]$. Therefore $C$ will be the same for the state $\tilde{\rho}(-a^*,-b^*)$. By the same argument as above, we can then study rather the mixture $\rho'=\demi\left[\tilde{\rho}(a,b)+\tilde{\rho}(-a^*,-b^*)\right]=\tilde{\rho}(iA,iB)$ with $A=\mathrm{Im}(a)$ and $B=\mathrm{Im}(b)$. This last state is Bell-diagonal:
\ba
\rho'_{AB}&=&\sum_{k=1}^4 \lambda_k\,\ket{\Phi_k}\bra{\Phi_k}
\ea where $\ket{\Phi_{1,2}}=\frac{1}{\sqrt{2}}\big(e^{i\chi}\ket{00}\pm e^{-i\chi}\ket{11}\big)$ and $\ket{\Phi_{3,4}}=\frac{1}{\sqrt{2}}\big(e^{i\chi'}\ket{01}\pm e^{-i\chi'}\ket{10}\big)$. The parameters are related as follows. Let $A'=\sqrt{(\mu_1-\mu_2)^2+A^2}$: then $\lambda_{1,2}=\demi(\mu_1+\mu_2\pm A')$ and $\cos^2\chi=\demi+(\mu_1-\mu_2)/A'$. The expressions of $\lambda_{3,4}$ and $\chi'$ are similar with $\mu_{3,4}$ and $B$. In particular, $C$ has the same value as above and now reads
\ba
C&=&2\left[(\lambda_1-\lambda_2)^2+(\lambda_3-\lambda_4)^2\right]\,.\label{Clambdas}
\ea

The \textit{third step} is now formally identical to the one for the BB84 protocol (we refer to Appendix A of \cite{sc-rmp-81-1301} for details). The four non-negative numbers $\lambda_j$ are constrained by three conditions: they must sum up to 1 and yield the measured values of $Q$ and $C$. This leaves one parameter free, that will be chosen as to maximize Eve's information. The first two constraints are taken into account by choosing the parametrization $\lambda_1=(1-Q)\frac{1+u}{2}$, $\lambda_2=(1-Q)\frac{1-u}{2}$, $\lambda_3=Q\frac{1+v}{2}$, $\lambda_4=Q\frac{1-v}{2}$, where $u,v\in[0,1]$; in which case, Eve's information reads
\ba
I_{E}(Q,u,v)&=&(1-Q)\,h\left(\frac{1+u}{2}\right)\,+\,Q\,h\left(\frac{1+v}{2}\right)
\ea where $h$ is binary entropy. The third constraint (\ref{Clambdas}) reads $C=2\left[(1-Q)^2u^2+Q^2v^2\right]$ and we have to compute $I_{E}(Q,C)=\max_{C} I_{E}(Q,u,v)$.

First note that $I_{E}(0,C)=h[(1+\sqrt{C/2})/2]$. For $Q>0$, we have $v(u)=\sqrt{C/2-(1-Q)^2u^2}/Q$; the condition $v\in[0,1]$ translates as $u\in [u_{min},u_{max}]$ where $u_{min}=\frac{1}{1-Q}\sqrt{\max[C/2-Q^2,0]}$ and $u_{max}=\min[\frac{1}{1-Q}\sqrt{C/2}\,,\,1]$. We have not found an analytical optimization for the whole parameter range.  However, $Q$ is expected to be small in a practical implementation; and for all $Q\lesssim 15.9\%$, one can show that $\frac{d}{du}I_{E}(Q,u,v(u))$ is strictly positive between $u_{min}$ and $u_{max}$, for all $C$; whence
\ba
I_{E}(Q,C)&=&I_{E}\left(Q,u_{max},v(u_{max})\right)\,. \label{evesinfo}
\ea
A rapid benchmark for qubit protocols is their robustness to white noise. For Werner states, $C=2(1-2Q)^2$: assuming this relation, we find $I_E(Q,C)=Q+(1-Q)h[(1-3Q/2)(1-Q)]$. This is exactly the same expression obtained for the six-state protocol \cite{lo01,sc-rmp-81-1301}. The corresponding secret key rate $r=1-h(Q)-I_E$ is positive for $Q\lesssim 12.62\%$, so well within the validity of (\ref{evesinfo}).\\

\begin{figure}[b]
\includegraphics[trim=0 250 190 200, clip, width=8cm]{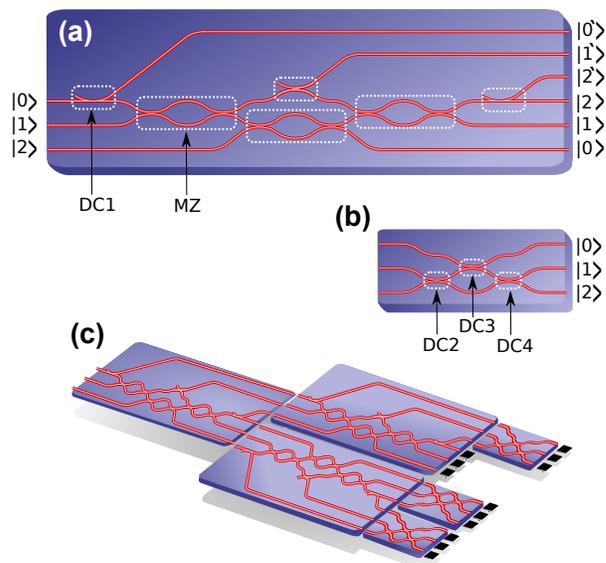}
  \caption{(Color Online) Integrated photonic components for measurement in the qutrit version of reference frame independent QKD. (a) The state splitter chip takes an arbitrary qutrit input state and splits it into a superposition of two probabilistic copies.  The reflectivity of directional couplers (DC) can be set to select the relative probability of the copy.  Two directional couplers implement a Mach Zender interferometer (MZ) with internal phase such that a photon exits from the path opposite to the one in which it entered. (b) A qutrit Hadamard chip takes a particular equal superposition basis and rotates to the computational basis in preparation for measurement. (c) Three state splitter chips are used to make a superposition of four probabilistic copies of the incoming states.  One probabilistic copy is immediately measured in the computational basis while the other three are fed into different Hadamard chips before measurement.
}
 \label{Figure 2}
\end{figure}

\textit{Extension to higher dimensions. ---} Several QKD protocols using higher-dimensional quantum systems (qudits) have been proposed, see e.g. \cite{cerf02}. In principle, they yield both higher key rates and larger robustness to noise. Qudit encoding in photonic states has been demonstrated using angular momentum modes \cite{angmom} or time-bins \cite{rob}. However, the control of the various relative phases (i.e. the stabilization of the reference frame) is very delicate: this is the reason why practical QKD has largely ignored higher-dimensional protocols. Even at the theoretical level, to our knowledge, nobody has explicitly computed security bounds against coherent attacks for these protocols, even if the general theoretical framework is in principle the same as for qubits.

A generalization of the rfi protocol, by removing the need for frame alignment, may provide the benefits of higher-dimensional encoding without the technical problems. Here we present such a generalization for qutrits. The derivation of rigorous security bounds for qudit protocols is a challenge in itself and is left for future work.

It is known that $d + 1$ sets of mutually unbiased bases (MUBs) exist for particles of dimension $d$, where $d$ is a power prime \cite{Iv-jpa-14-3241, Wo-aop-191-363}.  The joint space of any pair of qudits can be quantified by the $(d +1)\otimes(d + 1)$ measurements.  The protocol requires Alice and Bob to share an ensemble of qudit Bell states and randomly project their own particles onto the MUBs.  Their joint computational basis outcomes provide the $d$ dimensional key which is impervious to the effects of a changing phase between the computational states.  The joint outcomes of the complementary bases from the remaining $d^{2} + 2d$ measurements are used to calculate a fixed-but-unknown phase invariant quantity $C_{d}$, the higher dimensional analogue of the qubit case $C$.

For example, a natural operator representation of MUBs are the so called Weyl operators which have been studied in the context of entanglement \cite{be-jpa-41-235303, kl-pra-79-52101, ka-pra-80-22112}.  In the case of the $d = 3$ qutrit the Weyl matrices are often denoted by the set of eight $\tau_{i}$ matrices, each of which has a conjugate transpose twin in the set, with the same eigenvectors but with two permuted eigenvalues. $C_{3}$ is calculated on the unique eigenvector half set (neglecting the key forming computational basis operators).  With joint expectation values defined by $e_{ij}=Tr(\tau_{i}\otimes \tau_{j}.\rho_{AB})$ we find
\begin{equation}
C_{3} = \sum_{i,=2}^{4}\sum_{j,=2}^{4} e_{ij} e^{*}_{ij} + \sum_{i,=2}^{4}\sum_{j,=-2}^{-4} e_{ij} e^{*}_{ij} \leq 3
\end{equation}
where $\tau_{1}$ is the computational basis operator and those operators with negative indices are the conjugate transpose twin.  The maximal value is $C_{3}=3$, achievable only by (a subset of) two-qutrit maximally entangled states.

One possible physical implementation of the qutrit version of rfiQKD uses integrated photonic waveguides.  The rigid monolithic structure provides phase stability between spatial modes, so while chip-to-chip communication is phase-unstable, all unitaries implemented on-chip are highly stable.  With a network of variable beam splitters, or directional couplers (DCs), one can implement any unitary operator \cite{re-prl-73-58}.  A pair of maximally entangled qutrits can be created on a single chip via post selection and with the aid of ancilla photons \cite{kn-nat-409-46, ra-pra-65-012314}; alternatively one may use a spontaneous parametric down conversion source and select three pairs of points on the down conversion cone \cite{ce-prl-103-160401}.

To measure the incoming qutrits, Alice and Bob each require a device that randomly projects onto the four mutually unbiased bases.  This device may be assembled from two types of components: a state splitter and a qutrit Hadamard gate.  The state splitter is a $3$ input mode by $6$ output mode circuit that splits the incoming signal with three directional couplers of equal reflectivity and permutes the order of modes with Mach Zender interferometers, as shown in Fig 2a.  The qutrit Hadamard device, shown in Figure 2b, is composed of three directional couplers. In terms of Pauli matrices, $DC2 = \frac{1}{\sqrt{2}}(\sigma_{z} + \sigma_{x})_{12}$; $DC3 = \frac{1}{\sqrt{3}}(\sigma_{z} + \sqrt{2}\sigma_{x})_{01}$; $DC4$ = $\frac{1}{\sqrt{2}}(\sigma_{z} + \sigma_{y})_{12}$, where the modes acted upon are noted by the subscripts.  One can confirm that $DC4.DC3.DC2$ is a matrix in the Hadamard set; all other Hadamards in the set are accessible by adding phases to two of the three modes \cite{pr-pra-68-052315}. Three state splitters and three Hadamards fit together to make the random projector device shown in Fig 2c.

\textit{Conclusion. ---} We have described a protocol for exchange of a secure quantum key in an unknown and slowly varying reference frame and identified specific cases in which the protocol is useful.  More general scenarios can also be envisaged, for example, rfiQKD may be useful in an environment of intermittent rapid fluctuation where the key is exchanged during the periods of relative stability without the need to realign the reference frame.  We expect further situations in which rfiQKD is helpful to emerge.  We have provided a security proof for the qubit version of the protocol and described how the protocol can be developed into higher dimensions, with specific details of physical implementation in the qutrit case.\\

\textit{Acknowledgments. ---} This work was supported by EPSRC, QIP IRC, IARPA, ERC, the Leverhulme Trust, EU IP QAP (IST015848), the National Research Foundation and the Ministry of Education, Singapore. J.G.R. and J.L.OÕB acknowledge Royal Society Wolfson Merit Awards.

\bibliography{rfiQKD_al10}

\end{document}